\documentclass[aps,amsmath,twocolumn,floatfix,showpacs]{revtex4}
\usepackage{graphicx}

\begin{document} 

\title{A Mathematical Theory of Interpersonal Interactions \\ and Group Behavior}

\author{Yaneer Bar-Yam$^1$ and David Kantor$^2$}

\affiliation{$^1$New England Complex Systems Institute, Cambridge, Massachusetts 02142 \\
$^2$Kantor Institute, Cambridge, Massachusetts 02138}

\date{\today}

\begin{abstract}
Emergent collective group processes and capabilities have been studied through analysis of transactive memory, measures of group task performance, and group intelligence, among others. In their approach to collective behaviors, these approaches transcend traditional studies of group decision making that focus on how individual preferences combine through power relationships, social choice by voting, negotiation and game theory. Understanding more generally how individuals contribute to group effectiveness is important to a broad set of social challenges. Here we formalize a dynamic theory of interpersonal communications that classifies individual acts, sequences of actions, group behavioral patterns, and individuals engaged in group decision making.  Group decision making occurs through a sequence of communications that convey personal attitudes and preferences among members of the group. The resulting formalism is relevant to psychosocial behavior analysis, rules of order, organizational structures and personality types, as well as formalized systems such as social choice theory. More centrally, it provides a framework for quantifying and even anticipating the structure of informal dialog, allowing specific conversations to be coded and analyzed in relation to a quantitative model of the participating individuals and the parameters that govern their interactions. 
\end{abstract}


\maketitle

\section{Introduction}

Individuals interact to form couples, families, teams, organizations and societies. Understanding the shift from individuals to groups requires recognizing how collective behaviors arise and their properties. Understanding collective behaviors in many contexts is difficult due to the problem of observation of the processes involved \cite{dcs}. However, human groups are arguably the easiest complex system to observe as we see them all around us in professional and personal contexts. The challenge in this case is, at least in part, knowing what to look for and how to interpret what is observed. 

One of the exciting developments in understanding group collectivity and effectiveness has been the identification by Wegner \cite{wegner1987transactive} of the concept of transactive memory, a form of ``group mind" by which individuals who share experiences and know each other can recover memories more effectively than they can separately. Originally studied in couples, this concept has motivated wide ranging research on how it arises and its properties, including the use of transactive memory in professional teams \cite{hollingshead1998communication,hollingshead1998retrieval,brandon2004transactive,lewis2003measuring,lewis2004knowledge,alavi2002knowledge,moreland2006transactive,austin2003transactive,moreland2000exploring,reagans2005individual,mohammed2010metaphor,anand1998organizational,kanawattanachai2007impact}. 

The recovery of memories is one of many group behaviors that might be identified. A wider ranging body of research has considered a variety of group qualities linked to team task performance \cite{levi2015group,stewart1999team,thompson2008making,levine1998small}. While it makes intuitive sense that some level of individual ability is linked to a group's ability to perform certain tasks, other attributes have also been considered, including individual attributes associated to social behavior such as self-esteem, social sensitivity, social loafing, social status, and gender, and group processes such as goal setting, cohesion, trust, conflict, and sharing of communication time \cite{latane1979many,tziner1985effects,levine1990progress,karau1993social,o1994review,mullen1994relation,klein1995two,dirks1999effects,devine1999teams,duffy2000performance,duffy2002social,locke2002building,beal2003cohesion,de2003task,hong2004groups,kerr2004group,morris2005bringing,ilgen2005teams,van2007work,horwitz2007effects,lepine2008meta,van2008group,kleingeld2011effect,chiaburu2008peers,woolley2010evidence,shaw2011contingency,de2012paradox,waring2012cooperation,duffy2012social,de2014team,bates2017smart}. 

Another key group behavior that has been studied extensively is collective decision making. Two orthogonal mechanisms for group decisions are (1) assignment of decision making to an individual who makes decisions for the group as a whole, and (2) shared or collective decision making processes. The analysis of the former is through discussions of ``power" in many forms \cite{hobbesleviathan,john1959bases,galbraith1983anatomy,mann1984autonomous,french2001bases,keltner2003power,fiske2007social,michael1986sources,mann1993sources,mann2012sources3,michael2012sources4,baryam2018power}, while the latter has been extensively studied through analysis of voting systems (social choice theory) and negotiation. 

Decisions by voting and negotiation are generally considered to arise from what each individual would decide alone if given the decision authority (individual preference). Social choice theory considers how voting systems aggregate individual preferences with sometimes counterintuitive results  \cite{de2014essai,arrow2012social,gibbard1973manipulation,satterthwaite1975strategy,elster1989foundations,siegenfeld2018negative}. 
Negotiation, often defined as combining two or more divergent preferences into a joint agreement that must be unanimously accepted, originates in conflict in a zero-sum context  \cite{pruitt2013negotiation,lewicki2015negotiation,fisher2011getting}. 
It is often considered similar to or even an example of game theory \cite{von2007theory,myerson2013game,luce2012games} in which the decisions autonomously adopted by each individual have consequences for others and therefore decisions by individuals are linked. Actions involve exchanging information about preferences, making promises or threats.
Negotiation may, however, also include integrative (mutual benefit) considerations due to non-conflicted positive sum aspects of agreement or benefits of long term relationships \cite{pruitt2013negotiation,lewicki2015negotiation,fisher2011getting}. 

Another dimension of decision making in groups considers how one individual can influence another potentially resulting in fads and panics, and describing how common behavior and cultures arise \cite{bikhchandani1992theory,kindleberger2003manias,helbing2000simulating,harmon2015anticipating,Axelrod1997}. At the individual level, decision theory considers how individuals actually do or better should make decisions given individual values and uncertainty \cite{raiffa1974applied,slovic1977behavioral,einhorn1981behavioral,berger2013statistical,kahneman2013prospect,talebincerto}.

In this paper we are interested in how groups engage in decision making so as to benefit from combining individual capabilities beyond just their preference. For example, a group of individuals in a late afternoon meeting might decide to go out for dinner and where to go. While one person may have some of the information about what is a good decision, another may have additional information that modifies that decision. The best restaurant identified by one, might be closed due to renovation known to another. One knowledgeable individual may be reluctant to express their thoughts unless they are encouraged by another. The focus of this paper is on characterizing the sequence of communication among individuals that give rise to collective decision making in groups. Similar to transactive memory and group task performance as well as voting and negotiation we are not focused on the role of power itself, as in ``Who makes the decision," \cite{davis1976decision,solomon2014consumer} but on the way multiple individuals contribute to decision making. Recent discussions about debate in international relations point to several important distinctions in how people engage in collective decision making that are relevant \cite{risse2000let}.  In the first approach, rational actors with fixed preferences choose actions to optimize their utility in a social context, an essentially game theoretic approach. In the second, social interactions follow a set of rules and procedures embodying an assumption that mutual goals exist and are at least in part embodied by those social rules. Finally, actors also engage in meta communication about what assumptions are correct and what rules of social discourse apply in specific contexts. The latter opens the door to changes in individual opinions that affect the collective outcomes. 

Thus, collective decision making is not in general specified by rigid voting frameworks, and need not be either zero sum or reflect fixed and rigid individual priorities. Members of a group engaged in decisions need not be assumed to disagree. They may learn from each other. They may have values that are not entirely selfish, and may not even be focused on the particular decision at hand, considering other issues as equally or more important. For example, they may care more about the group cohesion or its decision making process. In the meta-view of their engagement in decisions, they may even consider each other to play a constructive part in a collective process of decision making that is shared. It is useful to consider examples that are mutual, like where to go for dinner, whether they may involve conflict at some point or not. Our interest is in the way interpersonal interactions give rise to collective decisions in such a context. The dynamics of communication arise from a much more fluid set of interactions than considered in voting. Kantor's theory of interpersonal interactions\cite{kantor1975inside,kantor2012reading} provides a framework for characterizing the set of communications that occur in such contexts. We formalize this framework within a mathematical theory of social collective behaviors. Compared to voting options in social choice theory, Kantor's theory considers a larger set of possible actions that combine to form a dynamic decision making process. 
However, the theory still identifies a limited number of categories of actions, providing a way to clarify the dynamics of  the large set of communications that can arise. The theory can be mapped onto real world decision making including both informal group discourse, and the more formal processes widely used, i.e. in Robert's rules of order \cite{robert1896pocket}. The complexity of the latter reflects the need to capture important details that are not embodied in social choice theory. 

In voting the available actions in a preset voting system are votes choosing among a predetermined set of options based on preferences, or abstention. This does not include the action by which the set of options are determined, an essential part of the process of decision making. Thus it is natural to include four possible options: move (propose a new action), follow (vote in the affirmative), oppose (vote in the negative), and by-stand (abstain). (We note that as in social choice theory sequential comparison of collective preferences may give rise to cyclical behaviors that do not converge to decisions. In expanding the set of options we make no a-priori judgement about whether the dynamics of the system is functional.) Kantor's theory posits that these four actions constitute the essential categories of communications in a group. In addition to distinguishing these four actions, there are different domains in which such actions can take place that are directly or indirectly relevant to the action taken. Kantor's theory presents three: power, meaning and affect. Others may be added in the fundamental description of the theory. Each domain can have each of the four possible actions. 

The basis of our contribution is to show that Kantor's theory is a universal characterization of the interactions within small groups of complex entities engaged in collective behavior with wide applicability to biological and social systems at all levels of organization. It is useful because it categorizes the many possible communications by individuals in a group to a small set that can be used to abstract communication patterns. The model is well suited to identifying when collective behaviors are functional or dysfunctional, for individuals or for collectives, in a context of environmental demands, individual and group needs. 

More technically, the universal process of decision making we describe is subject to the assumptions that (1) there is a high dimensional space of possible decisions, and (2) there are sequential communications of individuals to the entire group. The relevance of communication by each individual to the entire group is given by the connection between high dimensional potential decisions requiring a lot of information and the relevance of the entire group to the ultimate decision. This model is complementary to other models of decision making in which (a) all individuals make a synchronous contribution to the decision making as is described in the case of many voting systems, but also in the case of collective dynamics that is found in neural networks where the input (sensory information) is a high dimensional state and output (decision or action) is either low dimensional (as in pattern recognition in feedforward networks, which combines sequential and parallel communications) or high dimensional (as in memory retrieval in attractor networks) \cite{dcs}, and/or (b) there is a binary or few options of action. The latter is also consistent with a determination of a reduced dimensional set of options that is extrinsic to the process of decision making (this case is in part modeled by voting theory in which voting options are pre-specified). We will see that cases that are previously discussed in social choice theory arise as special cases of the asynchronous sequential dynamic decision making that we discuss here. Because of its sequentiality and high dimensional decision making the process we will describe is more relevant to small group decision making and so we consider it a universal process of small group decision making. A large group decision making process may be linked to small group decision making through the relevance of asynchronous communication to the small group that leads to a decision that is subsequently adopted by decision of a larger group through a synchronous decision making process (e.g. a vote). Or a hierarchical embedding of small within large or large within small group decisions can lead to hybrid processes. Thus a small number of large groups can interact sequentially overall and synchronously within, or a large collection of small groups can interact sequentially within and synchronously overall. Examples can be found in the real world. 

\section{Dynamic individuals and communications}

Individuals and groups have complex behaviors whose complete description would require high dimensional time histories. However, remarkably, if we restrict our attention to characterizing the relationships among actions of individuals, we can project the description to a few dimensions. The relationships among actions of individuals are precisely the dimensions that are the building blocks of the collective behaviors of the group. This reduced set of dimensions, as well as how individuals determine their actions within these dimensions, and the relationship of these actions to group behavior, is the core of the theory by Kantor. 

An individual can be quite generally characterized in an abstract fashion as a dynamical vector of attributes, $\psi_i(t)$, where the index $i$ indicates which individual. It is helpful to distinguish between slowly varying aspects, i.e. aspects that do not change much on a selected time scale of interpersonal interactions, e.g. one hour, and those which typically do change on that time scale or faster. Each of these rapidly changing attributes is modified in response to external stimuli that include the environmental conditions, and the internal dynamics of the individual. The internal dynamics include the consumption of energy as well as the mutual influences of neurons giving rise to the physiological basis of thought. The environmental conditions include the sensory impressions that arise from attributes of other individuals. 

\section{Collective movement: decision and follow-through}

We consider the role of interpersonal interactions to determine a direction of change of individuals relative to each other, and in particular to determine the pattern of collective action. Thus the nature of interactions is that they can point to, and the individuals can subsequently perform, movements that are coordinated. This enables individuals to act together as a group. 

To illustrate the process of collective interactions we consider a group of individuals in a late afternoon meeting discussing where to go for dinner. Given a set of possible options about places to eat, these individuals perform a set of interactions that result in them going together to a certain place for dinner, or the interactions may alternatively result in individuals going to different places for dinner. Either way, the framework of understanding the role of these interactions is to be considered in relation to the determination of where to go for dinner. The movement to dinner can be represented as a displacement of the attribute coordinates of the group of individuals, $\{\psi_i (t)\}$, which include as one attribute the spatial position of an individual. A displacement of all individuals from one original location to one final location constitutes a collective coherent motion which is one common pattern of collective behavior.

Generalizing this specific example, we can consider how individual actions aggregate to form collective actions. Quite generally, we argue that collective behaviors are formed when a set of individuals perform a set of interactions over a certain period of time. These interactions then determine a set of actions that result in attribute displacements that occur over a subsequent period of time. The dynamics can be considered as mathematically corresponding to the method of motion of certain bacteria through a liquid medium \cite{berg1972chemotaxis}. These bacteria perform a tumbling motion, followed by a straight line motion. This movement pattern can achieve rapid displacement while having the ability to direct the movement toward improved conditions such as presence of nutrients or absence of toxins. Since available information is local, the ultimate objective of movements is not known at the beginning of a set of movements, and episodic reassessment of the direction of motion is necessary. However, a continuous reassessment of direction of motion would limit progress much more severely. Thus a balance between frequency of assessment of direction, and period of motion is necessary to optimize the rate of progress, given properties of the environment such as the steepness of gradients of nutrients. Similar considerations apply to the decision making of groups of individuals and subsequent collective behaviors.

There are distinct attributes of the actions of individuals during the time of decision making and during the time of follow through activity. During the decision making period, individuals located together in a single space typically perform actions in sequence, one person at a time. This is due to the exclusivity of collective channels of communication in a single space (as well as the internal requirements of the sequential response dynamics described in the theory). During the follow through activity, generally individuals perform actions in parallel, unless the pattern of activity agreed upon specifically precludes it. Thus, during the decision about where to go for dinner, each person speaks in turn. During the period of walking to dinner, all are walking in parallel. The reason for the difference can be understood from the process of achieving consensus during the initial period, followed by the possibility of independent actions that are coherent in the second period. 

The requirements of the process of alternation of decision with action also implies a need to ensure that decision processes achieve closure through a time bounded process. This is essential even in the absence of perfect decision making or complete information. 

The example of people choosing to go to dinner is similar to the bacteria moving through a liquid medium in that both refer to spatial displacement. This should not be taken as a limitation. The same concepts can be applied quite generally to displacements in the very general attribute space having to do with the wide ranging set of possible activities engaged in by groups. 

\section{Discrete Communication Dynamics}

The attributes of multiple individuals over time affect each other during the period of decision making in a group. Actions by one individual that impact another individual are generally considered to be communications. While we can formulate the theory directly in terms of the general individual attributes, $\psi_i(t)$, it is helpful to start by simplifying the discussion by mapping these attributes onto communications, discrete units of activity that have shared meaning among individuals, given by an attribute vector $\xi_i(t) = M(\psi_i(t)).$ The purpose of this mapping is to enable us to consider two statements or gestures that may appear quite different, but share a common meaning, as having a similar description. Thus, if one person says ``Let's eat at  Sally's diner" and ``Let's go to the same place we had dinner at last night" map onto a similar meaning if Sally's was the place that dinner was eaten the previous night, even though the actual words spoken are different. Even if the words spoken are the same, their tonality, tempo and accent, varies among individuals, and for different instances by one individual. Similarly, it is possible that two statements have similar words but entirely different meanings, whether because of intonation (sarcasm) or that a single key word in the statement is different, i.e. insertion of the word ``not." Unlike $\psi_i(t)$, for $\xi_i(t)$ we can assume similarity in the attribute vector implies similarity of meaning in the domain of individual and group behavior. While specifics of the meaning map are interesting, and may play a role in our discussion, we bypass these details because it is tangential to the current purpose. Thus, for example, the issues that arise when meaning maps are not coincident among individuals are important but can be considered as an additional overlay to the theory we will develop.   

In particular we are interested in how one communication is related to previous communications:
\begin{equation}
\xi_i(t) = H[\{\xi_j (t')\}]
\end{equation}
where $\xi_i(t)$ is the communication of individual $i$, and $\{\xi_j (t')\}$ are the set of communications by individuals at prior times, and $H$ is a function that specifies how the current behavior is related to previous communications.

Such generalized models of dynamical behaviors have been used to describe the changing firing patterns of neurons in the brain, the development of color patterns on animal skins, the dynamics of panic in an auditorium, and other specific measures of individual and collective social attributes.\cite{dcs}  The main difference in the human communication model is the possibility of complex high dimensional communications, which while present in other cases is frequently not described as behaviors are abstracted as represented by single bits at a specific time. 

A specific statement may respond to only one of the previous communications. This would be written as: 
\begin{equation}
\xi_i(t) = H(\xi_{j(t)} (t'(t)))
\end{equation}
The choice of which statement to respond to adds to the complexity of the communications. It is possible for an individual to respond to the immediately preceding statement, or to a specific statement that subsequently collects multiple responses until a different statement becomes the statement that is responded to.  The individual who is speaking may choose which previous statement to respond to as well as choose the response. Alternatively, there may be a separate mechanism by which the statement to be responded to is determined, for example by rules of the group about ``motions" \cite{robert1896pocket}. This gives rise to a first form of meta communications, identifying what is the topic of conversation. More generally, statements may be a response to a collection of previous statements. For a first order analysis we consider only the case where an individual chooses a response to a specific previous statement, without rules for its determination. 

\section{Action categories: Move, Follow, Oppose, Bystand}

A key characteristic of a communication about a potential future action of the group (in brief, a conversational ``action") is its relationship to previous communications by others. We distinguish four categories of action relationships. A communicated action that is orthogonal to a previous action is labeled a ``Move", a parallel action is a ``Follow" an anti-parallel action is an ``Oppose" and a non-action is a ``Bystand". This is an exhaustive set of possible of high dimensional changes in a simple typology that is concerned with the process of individuals determining whether to move together and the direction to move in. In greater detail: 

\begin{itemize}
\item A significant action that is not related in a specific way to previous actions is an initiation of a direction of possible future movement of the group. This is labeled a ``Move". We can visualize it as a step of an individual in a direction of a particular place to eat, or the verbal analog of such a step, ``Let's eat at Sally's diner". 

Mathematically this corresponds to a vector that is orthogonal to previous moves, i.e. is in a new direction:
\begin{equation}
\begin{array}{ll}
\xi_i(t) \cdot \xi_j(t') &= 0 \\
| \xi_i(t) | &> 0
\end{array}
\end{equation}
The first equation specifies that there is no projection of the action on the directions of previous action by others. The second equation specifies that the action has significant magnitude. 

\item A significant action that is in the same direction as a previous action is labelled a ``Follow". It corresponds to a step consistent with participating in a coherent collective action.  

Mathematically this corresponds to a vector that is parallel to the previous move responded to:
\begin{equation}
\begin{array}{ll}
 \xi_i(t) \cdot  \xi_j(t') &> 0 \\
| \xi_i(t) | &> 0
\end{array}
\end{equation}

\item A significant action that is in the opposite direction of a previous action is labelled an ``Oppose". It corresponds to a step inconsistent with participating in a collective action in the original direction.  

Mathematically this corresponds to a vector that is anti-parallel to the previous move responded to:
\begin{equation}
\begin{array}{ll}
 \xi_i(t) \cdot  \xi_j(t') &< 0 \\
|  \xi_i(t) | &> 0
\end{array}
\end{equation}

\item Finally, a nonsignificant action corresponding to staying in place, is labeled a ``Bystand".  

Mathematically this corresponds to a vector of zero (small) magnitude:
\begin{equation}
\begin{array}{ll}
|  \xi_i(t) | &\approx 0
\end{array}
\end{equation}

\end{itemize}

We might further characterize the magnitudes of specific action overlaps, and a single action may have nonzero projections along multiple prior directions of action. These can be considered as additions to the basic typology given above. In the basic typology we are concerned with the ``sign" of the response by one individual to another, where sign in this case is a three valued function, taking the values $\pm 1$ as well as zero. 

\section{Synchronous or asynchronous communications}

Communication among individuals who are part of a group are to varying degrees parallel/synchronous or serial/asynchronous. Where a communication medium is exclusive at a particular time, the communications are properly modeled as serial. Where they are not exclusive they may be in parallel. Typically spoken auditory communication is considered to be exclusive, though it is possible for multiple individuals to speak at once---the difficulty with multiple parallel communications arises in the limitations on cognitive processing of the individual that receives multiple communications. Therefore, social norms more or less strictly limit spoken communications to be serial. Applause is another channel of auditory communication, which is considered inherently synchronous. Visual communication, i.e. facial expression or posture, might be considered to be synchronous because each individual has a posture or expression at all times, and one person can see multiple individual in their field of view. Still, expressions are often held for a period of time and changed at discrete times, whereby the changes of expression attract attention, and an individual generally only pays attention to one other individual at any particular time. Thus we see that these modalities are only approximately serial or parallel, but might be treated in one or the other way to first approximation. 

The differences in serial and parallel communication channels are also relevant to the types of communication that occur. Moves are generally limited to be serial. Each Move, being a high dimensional communication distinct from prior actions, must be communicated and separately processed by the individuals who receive it. Moreover, the responses to one Move must be separated from the responses to other actions, in order to be meaningfully related to group behavior.  On the other hand, Follows and Opposes can be highly parallel, as the amount of information communicated may be as simple as the fact of the follow or oppose, which can be communicated in a single bit of information. I.e., considering Follows and Opposes  together, the minimal communication in response to a Move is a binary response which is a vote of Follow or Oppose. Along with a Bystand, we have a single ``active" bit and a ``null" or abstain passive response. While this is the minimum amount of information in a response to a Move, this does not imply that every Follow or Oppose is restricted to this amount of information. Where such a reduced information is applicable, the aggregation of information can be done through a tally (a vote) considered as a representation of the group preference. From this we see that the conventional abstraction of group interactions into proposal and voting, is a simplified limit of the more complete group communication action model of Kantor. We also note that the result of a tally is itself a statement that is distinct from other statement types and can be responded to by individuals of the group. 

Typical small group Follow or Oppose, and even Bystand, communications have larger amounts of information than a single vote. Thus, the limitation of attention and processing by each individual restricts the number of parallel communications that can be effectively transmitted between individuals, especially when they are communicated among an entire group.  Thus, it is reasonable to consider a serial model of communication and we will focus on such a model here. (We note that a fully parallel model could also be constructed. In such a model, each individual at each time would respond to the other individual's actions at a previous time, or multiple previous times.  Such a model of response of each individual to the set of prior actions of the whole group, would be similar to a dynamic models of a recursive neural network, or recursive dynamic model of pattern forming of pigment cells.\cite{dcs} However, as mentioned previously, in such models each communication by a cell is generally represented as a single bit and this simplifies the structure of the model in a different way.)

Asynchronous models have a key additional feature that must be considered---the order of response. In many abstract models of asynchronous communication, the next responder is chosen at random. However, in a social group context, individuals may have reasons or character traits that result in actions in a certain order. Some individuals may and generally do act more frequently than others. Identifying the dynamics of sequence must be considered part of the description of the group dynamics. 

Thus, in an asynchronous model, parameters are needed to describe the degree to which an individual chooses to act, or not to act, i.e. their tendency toward inhibition. In the simplest model we will assume that there is a single parameter describing this tendency. Note that because a less reticent person will communicate before a more reticent person, and it is necessary at minimum to have two people to have a conversation, conversations may tend to be dominated by a few individuals, no matter how many people are present. 

Moreover, since individuals are not required to act in most groups, it is generally not possible to know what is the response of an individual unless he or she chooses to voice it. This creates an additional dynamic of withheld or hidden information that is integral to group behavior and its very nature makes it difficult to characterize. 
While there are other mechanisms that obscure communications, including their degree of truthfulness, we see that the existence of hidden information is part of group communications even in the presence of truthfulness.  

Indeed, the sequentiality and non-determinism of order give rise to a need for meta-communications about human group interactions. These meta-communications are directed at revealing hidden information, which may be key to group decision making, especially where exposing the participation or non participation of individuals in group behaviors under consideration is relevant to the ultimate outcome, and thus the consequences for the group. Sequentiality and meta-communications also give rise to the formal and informal structuring of conversations by ``rules of order." This second topic of rules, where the structure of communications are formalized, are typically designed so that more individuals participate and the conversation is compelled to achieve closure. Methods, include polling of individuals, and social conventions about required participation in collective behaviors for certain poll results. Without such structures, it may be impossible to determine what the collective decision is, as only some of the individuals may choose to participate in the communications, and typically only a few will. Rules of order, structure the time for decision making, and compel the group to remain together despite dissent. 

We see that social choice theory and the rules of order, can be considered founded on a more fundamental concept of group social interactions, which in a certain limit simplifies to those theories of group behavior. This suggests the underlying conceptual basis that we are developing can address important issues in group behavior. 

\section{Association dynamics}

The dynamics of group decision making is also inherently linked to the dynamics of group association, which characterizes the inclusion or exclusion of individuals, and the association or dissociation of the group as a whole. Among the relevant questions for our consideration are the coupling of group decision making to the group association, and the coupling of the behavior of individuals during group decision making to their individual choice to participate, or group choice to include or exclude them. Simply put, whether an individual's preferences are aligned with group decisions affects both the group's and the individual's desire to be part of the group. For couples, this dynamics include whether to marry or divorce. For business this includes incorporation and disbanding of associations. Group association dynamics can be refined to include strength of association or level of group engagement. 

A ``zeroth" model of group association would have any decision be itself a decision to associate or dissociate. Each individual would act according to his/her own statements, resulting in the final mover and followers going together, the opposers going in different directions, and bystanders staying in place. A first model of group association would have a single parameter characterizing group adhesiveness or ``surface tension," that tends to keep the group together. The strength of this force determines the degree to which opposers participate in the eventual group activity. Even a small amount of (positive) cohesiveness would imply bystanders participate. 

Group decision making may also directly engage in determining group composition. This is the third type of meta communication we have identified. In the discussion below, we will see that specific personalities may dominate this domain of discourse due to preferential reactions to different individuals. 

\section{Organizational structures}

The purpose of organizational structures is to habituate roles of individuals in social communications.  

Quite generally, groups organized around functions have mechanisms for achieving collective actions. These mechanisms include the dynamics of social interaction. In order to streamline the process of decision making so that the resources it consumes are reduced, the organization structure can formalize roles of individuals. The ``leader" engages in Moves, and others are followers engaging in Follows. Bystanders may exist to fulfill particular functions when needed. Consistent with the dynamics of progressive refinement of group association, individuals who tend to Oppose do not continue to be part of a group.

We will discuss later the functional benefits to the group of the dynamics of different roles under different circumstances. This will include the potential constructive role of opposers. Still, organizations that strive to include opposers, would tend to limit their roles in scope and effect.

\section{Dynamic model}

A mathematical model of group decision dynamics constructs the type of actions that are taken in terms of types of actions that have been taken in the past. In the simplest form, this would depend only on the most recent move (i.e. a deterministic iterative map or a stochastic Markov chain). More generally, we can consider a dynamic model that depends on a certain number of previous messages, the messages within a time interval, the most recent message of each of the individuals in the team, the most recent message of a particular type, etc. 

As a first approximation, it is conventional to consider a linear approximation of the dependency. We illustrate this for the case of a single communication depending on a single prior communication: 
\begin{equation}
\begin{array}{ll}
\xi_i(t) = h_i(t) + J_{i,j} \xi_j(t-1) 
\end{array}
\end{equation}
where $h_i(t)$ can be considered to be the intrinsic inclination of individual $i$, and $J_{i,j}$ is the (to first order fixed) linear response of individual $i$ to the comments of individual $j$. It is helpful to distinguish between three different cases for $J_{i,j}$. $J_{i,j}>0$ implies that $i$ tends to follow person $j$, $J_{i,j}<0$ implies that $i$ tends to oppose person $j$, and $J_{i,j} \approx 0$ implies $i$ tends to ignore person $j$. 

Testing the projection of the current on the previous state identifies the type of communication at time $t$ in terms of the communication at time $t-1$. 
\begin{equation}
\begin{array}{ll}
\xi_i(t) \cdot  \xi_j(t-1) &= h_i(t) \cdot  \xi_j(t-1) + J_{i,j} \xi_j(t-1) \xi_j(t-1) \\
&= h_i(t) \cdot  \xi_j(t-1) + J_{i,j} |\xi_j(t-1)|^2
\end{array}
\end{equation}
Considering only the sign $\pm,0$, i.e. the type of communication:
\begin{equation}
\begin{array}{ll}
\text{sign}(\xi_i(t) \cdot  \xi_j(t-1)) &=  \text{sign} (h_i(t) \cdot  \xi_j(t-1)) + J_{i,j}  |\xi_j(t-1)|^2)
\end{array}
\label{master}
\end{equation}
There are two different cases to consider:

The first case is if the comment by individual $j$ happens to have a projection, either positive or negative, along the direction of the intrinsic preference of individual $i$. In this case, we have a competition between the two terms. If the projection is positive and $i$ is inclined to follow $j$ the response will be a Follow. If the projection is negative, and $i$ is inclined to oppose $j$ the response will be an Oppose. If the projection is positive, and $i$ is inclined to oppose, or if the projection is negative, and $i$ is inclined to follow, the result will be determined by a balance between the amplitudes of each effect. 

The second case is if the comment by individual $j$ does not have a significant projection along the direction of the intrinsic desire of individual $i$. We note that because of the high dimensional nature of the possible communications, this can be expected to be the default case, i.e. a random choice of communication by individual $j$ would not have a projection along the intrinsic desires of individual $i$. (The magnitude of an inner product of two random vectors is the result of a random walk with a number of steps given by the number of vector components. The magnitude is proportional to the square root of the number of steps, which, for large number of components is small compared to inner products that scale as the number of components. \cite{dcs})
Only when the space of possible comments is for some reason limited, is it likely that the first case, where there is an intrinsic preference by the individual, applies. A simplest model would therefore set the first term to zero, eliminating the role of the intrinsic preference of the individual. In this case, we have the result that:
\begin{equation}
\begin{array}{ll}
\text{sign}(\xi_i(t) \cdot  \xi_j(t-1)) &=  \text{sign}(J_{i,j})  (1 - \delta(\xi_j(t-1),0))
\end{array}
\label{master_response}
\end{equation}
Neither factor depends on the specific content of the message. This equation has the immediate interpretation that to first (linear) order the projection onto the previous message, i.e. the category of the move, is independent of the message content. The response of one individual to a comment by another is then dependent only on the persistent relationship between the individuals. The final factor on the right of Eq. (\ref{master}) implies that to linear order a null communication has a null response. 

The analysis thus far focuses on the responses in terms of the prior communication that range from Follow to Oppose. A Move arises when the projection of $\xi_i(t)$ and $\xi_j(t-1)$ is small. This occurs for an individual for whom the influence of others is small ($J_{i,j} \approx 0$) or in a context in which the prior communication is itself of low magnitude $|\xi_j(t)|\approx0$. 
In this case the communication will be in the direction of the intrinsic desire of the individual, $h_i(t)$, with no significant projections on prior communications,  i.e. a Move. 

Finally, a Bystand occurs when an individual neither has a strong preference $h_i(t)$, nor a strong interaction $J_{i,j}$, or in case the two are balanced against each other. The case where both are small may also arise because of an inhibited response, which could be represented by an overall multiplicative factor that suppresses response. The inability, to first order, to observe the difference between inhibited responses and lack of preference is consistent with the actual challenge of understanding human response. The dynamics of sequentiality of acts must be considered carefully. People act or wait for a turn to act in the context of group interactions. Absence of a statement, maybe a wait state or may, under some conditions, constitute a Bystand --- i.e. does the person have an intended action and is waiting for their turn to speak, or are they choosing not to act. 

The model we have developed focuses our attention on the role of the intrinsic preference of the individual and the role of the intrinsic tendency of one individual in response to each of the other individuals independent of the content of the messages. 

\section{Power, Meaning, Affect}

When participants in a conversation make statements, their statements may be directed at moving the group to collective action directly, or may address aspects of cognitive, relational or other aspects of the system including meta-conversations about the subject, process of the dialog, or group dynamics. In general, a high dimensional attribute vector will have subdomains of attributes according to the modularity of the system. The nature of the modularity will vary among different kinds of systems. Here, some degree of non-universality enters into the discussion for application to different systems, though we expect that universality will still be significant. Statements can be categorized by what domain they are referring to. A different way to categorize the statements is by fundamental contributors of human response. For human social interactions, we identify three sub-domains of individual actions and their collective analogs. They are physical action oriented, cognitive meaning oriented, and affect oriented. From the point of view of the underlying vector of individual attributes, we can adopt conceptual or physiological languages to describe these different domains. 

Conceptually, the action oriented domain is associated to group behavior that manifests in physical activity, that of cognitive meaning is associated to changes in the interpretation of external stimuli or human communication, that of affect is associated with how individuals in the group feel about each other. We note that affect can be involved in how someone feels about a particular course of action, but the distinguishing characteristic of affect is its role in group relationships: ``I feel we should go together to Sally's diner" is an affect statement, while ``I feel we should go to Sally's diner" may not be. In terms of neurophysiological function, the action domain characterizes motor activity of motor neurons and muscles, the meaning domain characterizes cognitive processing, the affect domain characterizes the coupling of neural and endocrine systems. 

These three domains are identified by Kantor with the terms Power, Meaning and Affect. The use of the term ``Power" is suggested by the direct relevance of statements in this domain to action and when simply articulated are directives to action, though actual power over decisions may be shared due to statements made by others. The term ``Meaning" is suggested by the role of cognition in evaluating options for action. The term ``Affect" is linked to the distinct importance of emotional aspects of interactions to decision making by the group. 

Since communications among individuals are input through sensory channels, each of these domains is a characterization of part of the neural processing associated with response to environmental input, and may have associated output in the form of verbal or other communications. They are distinct in that, physical action is ultimately realized by non-verbal physical behavior, meaning is generally communicated verbally, affect is manifest through impacts on patterns of response and behavior. 

We will see from later analysis of the underlying mathematics of these interactions that the roles of these statements have implications for (a) directing action, (b) directing preferences about actions, (c) directing preferences about relationships among individuals. These essential distinctions suggest a degree of universality for these categories.

Each of the domains interacts with the other domains. For example, cognitive processing impacts on both actions of the motor system and the emotions associated to the endocrine system. The emotional state impacts on cognitive interpretation and actions. Actions taken impact emotional states and cognitive processing both through their external manifestations (including communications) and internal consequences. 

There are other aspects of discourse that may also be separated. One is that of values. Value statements potentially include a role in changing the group priorities that can affect all aspects of discourse including actions, preferences about actions, and preferences about relationships among individuals. In this paper we focus on Power, Meaning and Affect.

Mathematically, it is possible to write the vector of attributes of an individual, or the vector of attributes of the meaning of communications, as approximately partitioned into the three sub-domains, $\psi_i(t) = \{\psi^p_i(t),\psi^c_i(t),\psi^e_i(t)\}$ and 
$\xi_i(t) = \{\xi^p_i(t),\xi^c_i(t),\xi^e_i(t)\}$.

Specific models characterizing the ways each of these domains impacts on the other in terms of relevant mathematical equations can be constructed.  

For this paper it is more central to note that while it is intuitive to consider collective behaviors as characteristic of physical action, it is possible for the group to collectively perform actions that are primarily cognitive or affective. Correspondingly, the types of acts described earlier: Move, Follow, Oppose, and Bystand, can occur separately in each of these domains. Thus it is possible to perform a Move in the emotional or meaning domain as well as in the physical action domain. The partition of the domains is useful for the analysis of patterns of social interaction. In particular, it enables classifying a wide range of communications whose roles would not be apparent if the only types of collective action considered would be that of physical action. 

The interplay of Power, Meaning and Affect leads to a difficulty in disentangling them. By focusing on their primary roles, we can build a model that can be subsequently generalized to address other secondary roles that may be the most important under certain circumstances or for certain individuals. A first model of the dynamical roles of Power, Meaning and Affect is as follows: 

Power characterizes behavior that has consequences measured directly in terms of goals of the system---typically in this society these include health, fame and fortune, among others, and may vary in other contexts. An example is ``Let's eat at  Sally's diner," which can be understood as aimed at achieving a goal oriented activity for the group aimed at health, assuming that it is dinner time and people are hungry, and the food at Sally's diner is good.

Meaning actions are designed to impact on the members of the group and only indirectly on actions to achieve the goals of the group. In particular, they change the intrinsic desires of members of the group, $h_i$. An example of a Move in meaning space is ``I saw a cockroach at Sally's diner." The purpose of such a communication is two fold. First, so that the intrinsic preferences of individuals are modified so that subsequent actions quite generally and not just a specific one under discussion, are more likely to achieve goals (i.e. health). Second, so that the intrinsic preferences of the individuals of the group are more aligned with each other. Since a Move generally is in the direction of the intrinsic desire of an individual, this changes the analysis of the model in that $h_i(t)$ is partly aligned with $h_j(t)$. The result of such a Move is that future communications by one individual are more likely to be aligned with those of other individuals, leading to increased probability of collective behavior. 

Affect actions are designed to impact on the interpersonal relations within the group, and similar to meaning actions, they only indirectly impact on actions to achieve the goals of the group. In particular, they change the interaction matrix $J_{i,j}$. The central aspect of affect in this model is whether an individual wants to do something with another individual. The role of affect can be better understood when it is recognized that there are multiple satisfying patterns of social interaction that involve consistency among the actions taken. In particular, two individuals who both desire to be together, or both desire not to be together, can both be satisfied. On the other hand, two individuals are in a mutually incompatible situation if the first desires to be with the second, while the second does not desire to be with the first. Thus, desires of individuals can result in functional or dysfunctional states, depending on whether they are mutually consistent. Since the differences between these states are not inherently contained in objective conditions, they cannot be resolved by cognitive analysis. Affect enables individual actions to be selected based upon the self-consistency of relationships. Acts in affect promote a self-consistent pattern of mutual support or mutual antagonism, by adjusting the sign of $J_{i,j}$ to be equal to the sign of $J_{j,i}$. An example of a Move in affect space is ``I enjoy having dinner with you." The purpose is to communicate the positive sign of $J_{i,j}$ in order to prompt the second individual to recognize that mutuality is possible by a change in affect and therefore cause such a change in affect, i.e. a more positive $J_{j,i}$. Affect actions directly target mutuality and are inherently related to collectivity of a group, i.e. its tendency to act together or separately. Given that the high complexity of possible actions implies that most if not all Moves will be orthogonal to the intrinsic desires of others, as discussed before, the role of affect is critical in collective action.

In summary: Power actions are for achieving goals. Meaning is for aligning preferences with external / intrinsic consequences. Affect is for aligning desires of mutuality. We see that individual and collective movements can occur in these three domains, and can be analyzed in the same language. However, because each of these plays a different role in the individual, the significance of these individual and collective movements for the group are distinct. 

\section{Model parameters and their dynamics}

We consider individual inclinations $h_i(t)$ to be time dependent in the sense that an individual may get hungry and want to go to eat dinner. Other aspects of the inclination of an individual may be more persistent, such as causes a person is devoted to. Also, as previously discussed, communications in Meaning can affect the individual inclinations. 

We consider $J_{i,j}$ to be time dependent. A reasonable first model for such changes is that the presence of communications that are aligned (anti-aligned) of two individuals tends to make their $J_{i,j}$ more positive (more negative) respectively. This corresponds to mathematical models of Hebbian imprinting in neural learning where the individuals are analogs of neurons. This form of learning reinforces patterns of social interaction so that certain individuals tend to become more positive or negative relative to each other, causing those who tend to act in concert to be more likely to act in concert in the future, and similarly those who act in opposition will be more likely to act in opposition in the future., It also tends to make those who have compatible preferences tend also to have mutually reinforcing interpersonal relationships. Also, as previously discussed, communications in Affect can affect the individual inclinations. 

\section{Classification of individuals}

Different individuals tend to adopt different types of actions, which leads to a means for classifying them.  The reason for these choices are embedded in persistent properties of individuals rather than dynamic properties, which are therefore to be identified as personality types. We can identify the types of actions preferred by individuals from the categories of ``Move, Follow, Oppose, Bystand." 

In the mathematical model, an individual is identified by the parameters $h_i$ and $J_{i,j}$ for a particular value of $i$. These parameters generally vary for different values of the index, $j$, i.e. in relation to different people, and at different times. We identify personalities by postulating that for a particular individual these parameters have characteristic values that persist over time and satisfy identifiable patterns across people. 

The identification of those who frequently perform just one of the types of actions can be readily identified:

A person with strong preferences $h_i$ and weak interactions $J_{i,j}$ is goal/content driven, and frequently Moves. 

A person who has strong interpersonal interactions $J_{i,j}$ and weak preferences $h_i$ will respond to other moves. They Follow, if $J_{i,j}$ tends to be positive, or they Oppose, if $J_{i,j}$ tends to be negative. 

A person with weak preferences and weak interactions tends to Bystand. As discussed previously, an additional overall amplitude parameter can be considered as describing an inhibition of response. Thus, a persistent bystander may have externally unexpressed preferences and interactions. 

Among other insights, we see that the those who are frequent movers, often identified as ``leaders," have a tendency toward more limited social interactions compared to those who are followers and opposers.

Certain combinations of parameters also tend to generate more complex response patterns, specifically when the tendency is to adopt more than one type of action according to specifics of the circumstances. A person with strong preferences and strong interactions has a varied response. Such a person, in order to reduce conflicts between preferences and interactions may also tend to engage in Meaning acts, through whose adoption by the group, the conflict will be reduced. A person with weak preferences and strong interactions that vary significantly among individuals, i.e. with strong interpersonal preferences, will tend to follow some and oppose others. In order to reduce conflicts, between following and opposing individuals, such a person may engage in Affective acts to reduce conflict. Alternatively, in some sense equivalently, such an individual may act in the dynamics of selecting group membership.

\section{Summary}

The understanding of group interactions related to decision making may be advanced by considering a limited typography of communication acts. We have shown that it is possible from general mathematical considerations to identify four types of actions, $\{$Move, Follow, Oppose, Bystand$\}$, as universal, and using simple mathematical dynamic models to advance the understanding of group dynamics. We have explored some of the implications including identifying personality types and the relationship between types of actions taken and the types of personalities that are present as described by the model. The model we obtained follows closely the discussion and analysis of conversations in families and businesses by Kantor \cite{kantor1975inside,kantor2012reading}. The usefulness of this model in diagnosing functional and dysfunctional behaviors and interventions has already been demonstrated.

\end{document}